\documentclass[reprint,aps,prd,nofootinbib,preprintnumbers,twocolumns,showpacs,10pt, superscriptaddress]{revtex4-1}
\usepackage{epsfig}
\usepackage{graphicx}
\usepackage{amsmath}
\usepackage{amsfonts}
\usepackage{amssymb}
\usepackage{bm}
\usepackage[usenames,dvipsnames]{color}
\usepackage{hyperref}
\hypersetup{
    colorlinks=true,       
    linkcolor=NavyBlue,          
    citecolor=Magenta}

\newcommand\ee{\end{equation}}
\newcommand\be{\begin{equation}}
\newcommand\eea{\end{eqnarray}}
\newcommand\bea{\begin{eqnarray}}

\newcommand\mpc{\,\mbox{Mpc}}

\newcommand\ie{{\it i.e.}~}
\newcommand\eg{{\it e.g.}~}
\newcommand\del{\partial}
\newcommand\eq[1]{Eq.~(\ref{#1})}

\renewcommand\({\left(}
\renewcommand\){\right)}
\renewcommand\[{\left[}
\renewcommand\]{\right]}

\begin{document}
\preprint{IFIC/14-49}

\def\thefootnote{\fnsymbol{footnote}}
\title{On the current status of Modified Gravity}

\author{Lotfi Boubekeur}
\affiliation{ Instituto de F\'isica Corpuscular (IFIC), CSIC-Universitat de Valencia,\\
Apartado de Correos 22085,  E-46071, Spain.}
\affiliation{ Laboratoire de Physique Math\'ematique et Subatomique (LPMS)\\
Universit\'e de Constantine I, Constantine 25000, Algeria.}

\author{Elena Giusarma}
\affiliation{Physics Department and INFN, Universit\`a di Roma ``La Sapienza'', Ple Aldo Moro 2, 00185, Rome, Italy}

\author{Olga Mena} 
\affiliation{
Instituto de F\'isica Corpuscular (IFIC), CSIC-Universitat de Valencia,\\ 
Apartado de Correos 22085,  E-46071, Spain.}

\author{H\'ector Ram\'irez}
\affiliation{ Instituto de F\'isica Corpuscular (IFIC), CSIC-Universitat de Valencia,\\
Apartado de Correos 22085,  E-46071, Spain.}

\begin{abstract}
We revisit the cosmological viability of the Hu $\&$ Sawicki modified gravity scenario. The impact of such a modification on the different cosmological observables, including gravitational waves, is carefully described. The most recent cosmological data, as well as constraints on the relationship between the clustering parameter $\sigma_8$ and the current matter mass-energy density $\Omega_m$ from cluster number counts and weak lensing tomography, are considered in our numerical calculations. The strongest bound we find is $|f_{R0}| < 3.7 \times 10^{-6}$ at $95\%$~CL. Forthcoming cluster surveys covering 10,000 deg$^2$ in the sky, with galaxy surface densities of  $\mathcal{O}(10)$~arcmin$^{-2}$ could improve the precision in the $\sigma_8$-$\Omega_m$ relationship, tightening the above constraint.
\end{abstract}

\maketitle

\section{Introduction}

Current cosmological observations have robustly confirmed that the universe is expanding at an accelerating pace~\cite{planck,supernovae}. Unraveling the nature of the physics responsible for such a phenomenon is, together with the dark matter puzzle and the generation of the primordial perturbations which seeded the observed structures in our universe,  one of the most important problems in modern cosmology. \\

Understanding such accelerated expansion might lead to a revolution in our knowledge of particles and/or fields. The simplest and most minimal  explanation is to ascribe it to the presence of a vacuum energy density, \ie a {\emph cosmological constant} $\Lambda$, dominating the energy budget of the universe.  This is the so-called $\Lambda$CDM scenario, which corresponds to a flat universe made up of roughly $25\%$ of dark matter, 5$\%$ normal matter, and $70\%$ vacuum energy. This simple and economical scenario is in spectacular   agreement with existing observations, however it is extremely fine-tuned~\cite{Weinberg:1988cp,Copeland:2006wr}. An alternative possibility is to allow for a more general fluid, a \emph{dark energy} fluid, with an equation of state $w$ different from that of a cosmological constant, \ie  $w\neq-1$. Unlike in the $\Lambda$CDM scenario, this parameter might change during the expansion history. Scalar field models, in which dark energy is identified with a \emph{quintessence} scalar field~\cite{Caldwell:1997ii,Zlatev:1998tr,Wang:1999fa,Wetterich:1994bg}, are also a possible option to explain the current accelerated expansion. However, these models have  the same fine-tuning problem of the cosmological constant, as there is no symmetry in nature to explain the tiny value of the minimum in the quintessential potential.\\

The third possibility, which is the one explored in this work, relies on infrared modifications of the gravitational sector.~\emph{Modified gravity} involves a modification of Einstein equations on large scales, incorporating a plethora of models with either extra spatial dimensions or with an action which is non-linear in the curvature scalar (the so-called $f(R)$ theories,~\cite{Carroll:2003wy}) or higher order curvature invariants~\cite{Carroll:2004de}, see Ref.~\cite{DeFelice:2010aj} for a complete review. Among all the possible models, we shall focus here on $f(R)$ theories. Early work on the observational consequences of these theories concluded that they are compatible with both universe's background measurements~\cite{Mena:2005ta} as well with data involving perturbed quantities~\cite{Zhang:2005vt}. However, these models may not satisfy solar system constraints, because of the presence of an additional scalar degree of freedom, $f_R\equiv d f/d R$, that  mediates a long-range fifth-force~\cite{Chiba:2003ir}. Hu $\&$ Sawicki proposed a $f(R)$ model which evades solar system constraints, while mimicking $\Lambda$CDM at late times~\cite{Hu:2007nk}~\footnote{For the required conditions in the context of chameleon theories to have a cosmological impact, se Ref.~\cite{Wang:2012kj}.} In the following, we shall restrict ourselves to this model and to its observational signatures. 
Current cosmological constraints as well as the expected errors from future surveys on the parameters governing this model have been previously addressed by a number of authors~\cite{Martinelli:2009ek,Martinelli:2011wi,Marchini:2013lpp,Marchini:2013oya,Hu:2013aqa,Munshi:2014tua,Bel:2014awa}. Here we review the Hu $\&$ Sawicki model, paying special attention to the $B$-mode signal in light of the recent BICEP2 data~\cite{bicep2,bicep22}, as in Refs.~\cite{Amendola:2014wma,Zhou:2014fva}, where $f(R)$ theories have been analysed considering the impact on the gravitational wave spectrum.\\

The structure of the paper is as follows. We start in Sec.~\ref{sec:HS} with a brief introduction of the Hu $\&$ Sawicki model, describing its general behaviour, limits, and most important features. Section~\ref{sec:impact} focuses on the impact of the model on the different cosmological observables, such as on the Cosmic Microwave Background (CMB) Integrated Sachs-Wolfe effect, lensing and gravitational wave signals, as well as on cluster number counts and on weak lensing tomography. In Sec.~\ref{sec:data}, we describe the method and the data used in the cosmological analyses. The constraints arising from these analyses are presented in Sec.~\ref{sec:results}. In Sec.\ref{sec:conclusions}, we present our conclusions.

\section{The Hu $\&$ Sawicki  model}
\label{sec:HS}

The Hu $\&$ Sawicki model is described by the action
\begin{equation}\label{eq:action}
S=\int{d^4x \sqrt{-g} \big[\frac{R+f(R)}{2\kappa^2}+{\cal L}_{m}\big]}~.
\end{equation}
In the above equation, ${\cal L}_m$ refers to the matter Lagrangian, $\kappa^2=8\pi G_N$ and the $f(R)$ is given in terms of the Ricci scalar $R$  by 
\begin{equation}\label{eq:fr}
f(R)=-m^2\frac{c_1\(\frac{R}{m^2}\)^n}{1+c_2\(\frac{R}{m^2}\)^n}~.
\end{equation}
\noindent The parameter $m^2=\kappa^2\rho_0/3$, where $\rho_0$ is the  current mean mass-energy density and the parameters 
$c_1$, $c_2$ and $n$ are free parameters, to be determined by the observational constraints. These three parameters can be 
related to the effective relative matter ($\tilde{\Omega}_m$) and dark energy densities ($\tilde{\Omega}_x$) as  
\begin{equation}
\frac{c_1}{c_2}\approx6\frac{\tilde{\Omega}_x}{\tilde{\Omega}_m}~;
\label{eq:limit0}
\end{equation}
and
\begin{equation}
\frac{c_1}{c_2^2}=-\frac{f_{R_0}}{n}\(\frac{12}{\tilde{\Omega}_m}-9\)^{n+1}~.
\label{eq:limit}
\end{equation}
We assume in the following a flat geometry,  and therefore  $\tilde{\Omega}_x=1-\tilde{\Omega}_m$. 
Notice that the introduction of an effective dark energy component with an energy density given by $\tilde{\Omega}_x$ represents only a redefinition of the parameters, i.e. there is no a dark energy extra component in the universe, as the accelerated expansion is exclusively arising from the gravitational sector.

The parameters $c_1$ and $c_2$ are related to the parameters of the Hu $\&$ Sawicki model, $n$ and $f_{R0}\equiv f_R (\ln a =0)$, and to the effective matter energy density $\tilde{\Omega}_m$, as follows: 
\begin{eqnarray}
c_2&=&-6 \frac{\tilde{\Omega}_x}{\tilde{\Omega}_m} \frac{n}{f_{R0}}\(\frac{12}{\tilde{\Omega}_m}-9\)^{-n-1}~, \nonumber \\
c_1&\approx& 6\frac{\tilde{\Omega}_x}{\tilde{\Omega}_m} c_2~.
\end{eqnarray}

Notice that when $n\rightarrow 0$, $f(R)\rightarrow 0$, and therefore one does not recover standard $\Lambda$CDM cosmology, since there is no extra contribution from the gravitational sector to mimic the vacuum energy.  In the limit in which the curvature $R\gg m^2$, $f(R)$ reads:
\begin{equation}
 f(R) \approx -\frac{c_1}{c_2} m^2 +\frac{c_1}{c^2_2} m^2 \left(\frac{m^2}{R}\right)^n~,
\end{equation}
from which we can learn that the $\Lambda$CDM limit is recovered for either $n\rightarrow \infty$ and/or $f_{R0}\rightarrow 0$. As we shall see in Sec.~\ref{sec:results}, this is the parameter region preferred by current cosmological data. 

The modified Einstein equations for a generic $f(R)$ model are obtained after varying the action, see Eq.~(\ref{eq:action}), with respect to the metric $g^{\mu\nu}$:
\begin{equation}
G_{\mu\nu}+f_RR_{\mu\nu}-\[\frac{f}{2}-\Box f_R\]g_{\mu\nu}-\nabla_\mu
\nabla_\nu f_R=\kappa^2T_{\mu\nu}~,
\end{equation}
in which $f_{RR}\equiv d^2f/dR^2$. The resulting Friedmann equation is no longer algebraic but a second order differential equation, which reads:
\begin{equation}  
H^2-f_R\(HH'+H^2\)+\frac{f}{6}+H^2f_{RR}R'=\frac{\kappa^2\rho}{3}~,
\label{eq:fri}
\end{equation}
\noindent where the prime denotes derivative with respect to $\ln a$ and $H$ is the Hubble parameter. In order to solve the modified Friedmann equation, we follow the treatment of Refs.~\cite{Martinelli:2009ek,Martinelli:2011wi}, where two new auxiliary variables $y_H=(H^2/m^2)-a^{-3}$ and $y_R=(R/m^2)-3a^{-3}$ are introduced. It follows that Eq.~(\ref{eq:fri}) can be expressed as 
\bea
y'_H&=&\frac{y_R}{3}-4y_H~,\\
y'_R&=&9a^{-3}-\frac{1}{y_H+a^{-3}}\frac{1}{m^2f_{RR}}\nonumber\\ 
&\times&\[y_H-f_R\(\frac{y_R}{6}-y_H-\frac{a^{-3}}{2}\)+\frac{f}{6m^2}\]~.
\eea
After solving the background equations in terms of the new variables $y_H=(H^2/m^2)-a^{-3}$ and $y_R=(R/m^2)-3a^{-3}$, we obtain the Hubble parameter in the Hu $\&$ Sawicki model: 
\begin{equation}\label{eq:H}
H(a)=\sqrt{\tilde{\Omega}_mH_0^2(y_H+a^{-3})}~.
\end{equation}

To analyse linear perturbations, we shall adopt the conformal Newtonian gauge, where the line element reads 
\be
ds^2=-(1+2\psi)\,dt^2+a^2(t)\[(1-2\phi)\delta_{ij}+h_{ij}\]dx^i dx^j\,, 
\ee
where $\phi$ and $\psi$ are the usual metric perturbations variables  and $h_{ij}$ is the physical tensor mode satisfying $h_{ii}=\del^i h_{ij}=0$ (traceless and transverse). The $i-j$ $(i\neq j)$ component of the Einstein equations reads as~\cite{Bean:2006up}
\begin{widetext}
\begin{equation}
\label{eq:stress}
(1+f_R) (\psi-\phi)+2 f_{RR} \[-12H^2\psi -6\dot{H}\psi -3H \dot{\psi}- 12 H \dot{\phi} -3\ddot{\phi} +\frac{k^2}{a^2} \psi-2  \frac{k^2}{a^2} \phi\]=-8\pi G_N a^2 \Pi~,
\end{equation}

\end{widetext}

\noindent where the dot refers to derivative with respect to the time variable $t$ and $\Pi$ is the anisotropic stress tensor, present for non-perfect fluids, satisfying $\Pi_{ii}=\del^i\Pi_{ij}=0$ (transverse and traceless), which includes the anisotropic stress of both photons and neutrinos. Notice that, neglecting these lasts, the ratio of the metric perturbations is no longer $1$. As we shall see in the following, this gravitational slip between the metric perturbations is a very distinctive feature of modified gravity scenarios, which will help in distinguishing these models from other dark energy schemes.

\section{Modified gravity and Cosmological Observables}
\label{sec:impact}

\begin{figure}
\hspace*{-1.0cm} 
\vspace*{-0.cm} 
\includegraphics[width=10.5cm]{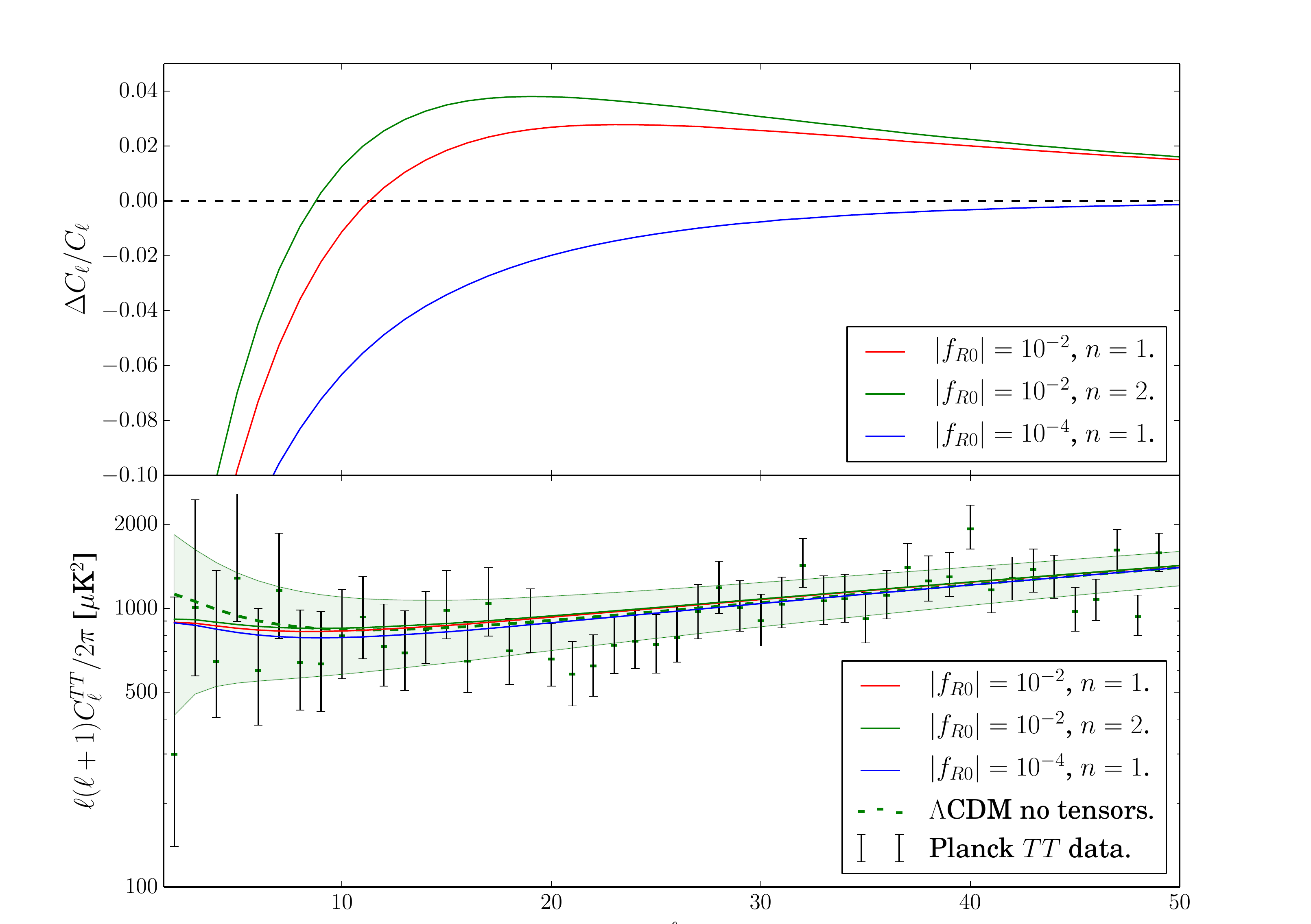}
 \caption{Top panel: Difference in the low $\ell$ CMB signal (relative to the corresponding $\Lambda$CDM model), for several possible values of the Hu $\&$ Sawicki  gravity parameters, and neglecting the tensor contribution. Bottom panel: the coloured solid lines refer to the ISW effect for the same models depicted in the top panel, together with the $\Lambda$CDM model prediction (dashed lines), with no tensor contribution. The light green (shaded) region depicts the cosmic variance uncertainty, while the data points are Planck low-$\ell$ temperature measurements~\cite{planck}.}
\label{fig:tensors1}
\end{figure}

\begin{figure}[t!]
\includegraphics[width=9cm]{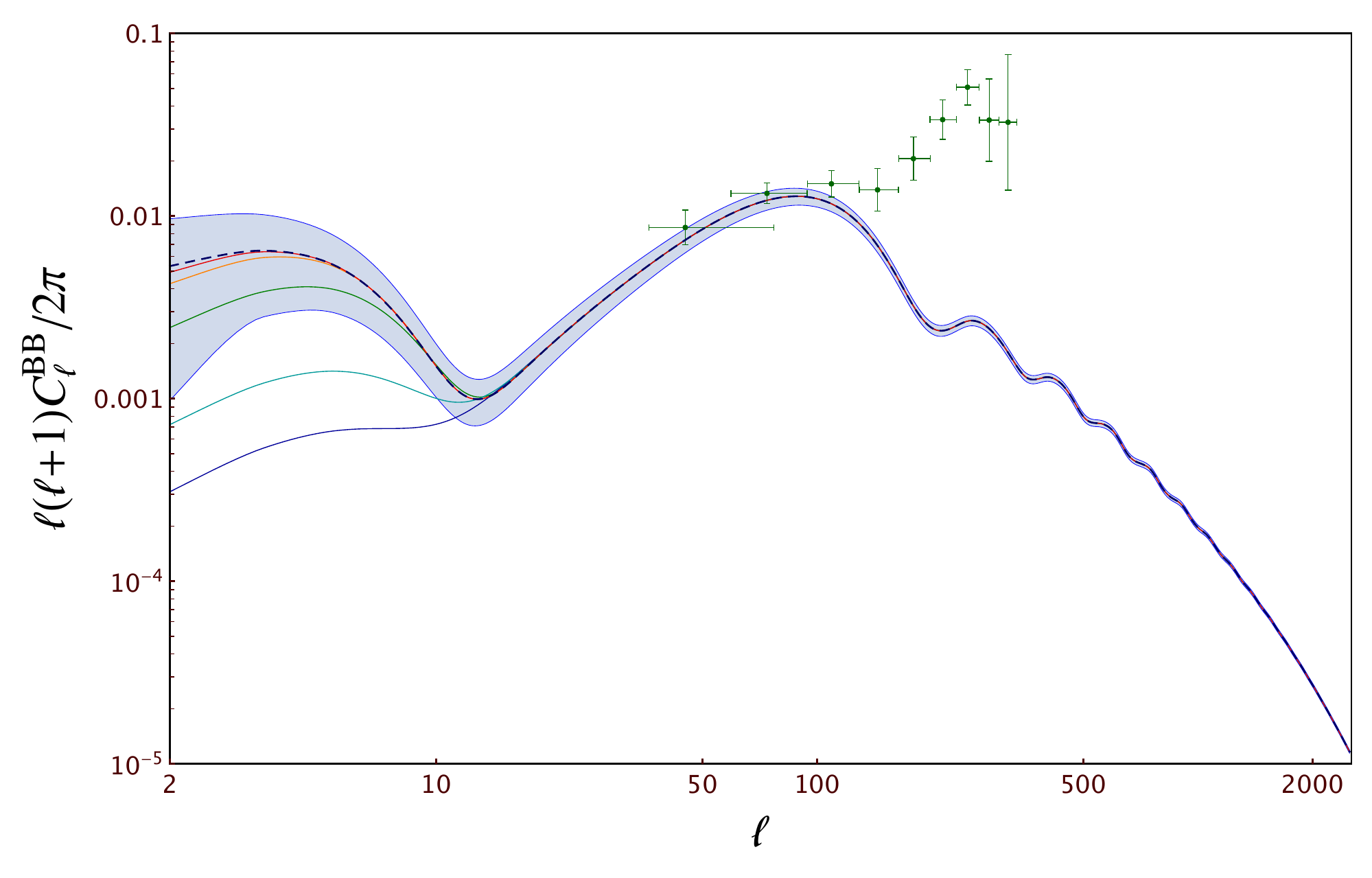}\\
 \caption{The solid lines denote the tensor power spectrum for the Hu $\&$ Sawicki model for several possible values of $a_{\textrm{f(R)}}=0.03,0.05,0.1, 0.2$ and $0.3$ (from the bottom to the top) and a tensor-to-scalar ratio $r=0.2$. The dashed lines refer to the $\Lambda$CDM model. The shaded region depicts the cosmic variance uncertainty. The data points are the recent BICEP2 measurements~\cite{bicep2}.}
\label{fig:tensors2}
\end{figure}
\subsection{CMB: ISW effect and lensing}

The Integrated Sachs Wolfe (ISW) effect~\cite{Sachs:1967er} arises from the time variation of the gravitational potentials. The corresponding additional contribution to the CMB temperature anisotropy $\left(\Delta T/ T\right)_{\textrm{ISW}}$ measures the time-dependence of the gravitational potentials along the line-of-sight, and is therefore proportional to $(\dot{\psi}+\dot{\phi})$. Modified gravity models can lead to a non-negligible signature in the ISW effect at late times, when the effects of the acceleration of the universe are important~\footnote{In the following, unless otherwise stated, we shall assume that Hu $\&$ Sawicki gravity starts operating at a scale factor  $a_{\textrm{f(R)}}=0.1$}. The non-zero anisotropic stress caused by the new terms arising in $f(R)$ theories becomes then significant, leading to a modification of the temperature anisotropies at low multipoles, see Eq.~(\ref{eq:stress}). The ISW effect in the context of Hu $\&$ Sawicki gravity has been explored by a number of authors, see Refs.~\cite{Song:2006ej,Hu:2012td,Cai:2013toa,Munshi:2014tua}. In the bottom panel of Fig.~\ref{fig:tensors1}, we plot the ISW effect for several possible values of the parameters of the model, together with the $\Lambda$CDM prediction (in dashed lines), the cosmic variance uncertainty and the Planck low-$\ell$ temperature data~\cite{planck}. The largest departure from the canonical $\Lambda$CDM predictions appears at large values of $|f_{R0}| \sim 10^{-2}$.  In general, the ISW effect decreases monotonically as $|f_{R0}|$ increases, showing a mild sensitivity to the value of $n$, in agreement with the findings of Ref.~\cite{Hu:2012td}. The top panel of Fig.~\ref{fig:tensors1} depicts the relative difference with respect to standard $\Lambda$CDM predictions, of the ISW effect in the  Hu $\&$ Sawicki scenario, for the same combination of parameters of the bottom panel. As previously stated, there is always a suppression of power due to the different ISW effect in this family of modified gravity models. 

CMB lensing measurements are, together with the ISW effect, another powerful CMB probe to test the nature of the current universe's accelerated expansion. It was soon realised that the CMB lensing signal will also be affected by the non-vanishing anisotropic stress that naturally appears in $f(R)$ theories~\cite{Amendola:2007rr,Acquaviva:2004fv}, as the lensing potential is the sum of the two metric perturbations $\Phi=\psi +\phi$~\cite{Lewis:2006fu}. In $f(R)$ theories, the lensing potential will therefore depend on an anisotropic stress which arises exclusively from the gravitational sector. More concretely, in the quasi-static approximation, for sub-horizon modes, the CMB lensed power spectrum will be modified with an additional factor which scales like the square of the gravitational slip between the metric perturbations (see \eg  \cite{Amendola:2014wma} for a recent treatment). We shall include CMB lensing in our numerical analyses to maximise the sensitivity of Planck data to modifications of gravity.

\subsection{Effect of modified gravity on CMB tensor modes}

In standard general relativity, Einstein equations for the tensor perturbations are given by  
\be
\ddot{h}_{ij}+3 H\dot{h}_{ij}+\({k^2\over a^2}\) h_{ij}=16\pi G_N\Pi_{ij}~,
\label{eq1}
\ee
where $\Pi_{ij}$ is the anisotropic stress tensor and we have followed the notation of Ref.~\cite{Weinberg:2003ur}. In modified gravity models, as previously stated, there is an anisotropic stress contribution from the gravitational sector, see Eq.~(\ref{eq:stress}). Such anisotropic stress has been extensively explored in the literature via different parameterisations~\cite{mganisotropic,Kunz:2006ca} \footnote{Modified gravity models are not the only ones inducing a non-zero anisotropic stress. There are also some possible dark energy models, with non-standard fluids, in which an additional contribution to the total anisotropic stress may arise~\cite{Kunz:2006ca,Bento:2002ps,deanisotropic}.}. The effect of such an anisotropic stress in the $B$-mode lensed signal for a particular case of modified gravity has been recently explored by the authors of  Ref.~\cite{Amendola:2014wma}.

Here, we follow a different approach: rather than parameterising the anisotropic stress induced by the modified gravity model explored here (as in Ref.~\cite{Amendola:2014wma}), we derive the evolution equations for  tensor perturbations in the presence of a generic $f(R)$ (see also Refs.~\cite {DeFelice:2010aj,Hwang:1996xh}):
\be
\ddot{h}_{ij}+{1\over (1+f_R) a^3}\del_t\[a^3  (1+f_R)\] \,\dot{h}_{ij}+{k^2\over a^2} h_{ij}=\frac{16\pi G_N \Pi_{ij}}{(1+f_R)}~.
\ee
Using the expression of Ricci scalar, $R=6(2 H^2 +\dot{H})$, the tensor perturbation equations in $f(R)$ theories reads 
\begin{widetext}
\be
\ddot{h}_{ij}+ \[3 H + 6(4 \dot{H} H+\ddot{H})\frac{f_{RR} }{(1+f_R)}\]\dot{h}_{ij}+\({k^2\over a^2}\) h_{ij}=\frac{16\pi G_N \Pi_{ij}}{(1+f_R)}~.
\label{eq:tensor}
\ee
\end{widetext}
Notice that in $f(R)$ gravity scenarios, the tensor equations are modified, containing a new damping term and an additional factor in  the source term. Since $f(R)$ gravity models have an impact on the universe's evolution solely at late times, when the neutrino and radiation energy densities are negligible compared to the rest of the universe's components, the modification introduced by $f(R)$ theories in the tensor equation source term turns out to be negligible. However, the additional damping term in Eq.~(\ref{eq:tensor}) can affect the tensor perturbation evolution, see Fig.~\ref{fig:tensors2} for an illustration of this effect on the predicted $B$-mode spectrum, for a tensor-to-scalar ratio $r=0.2$ and several possible values of the scale at which modifications of gravity switch on, $a_{\textrm{f(R)}}=0.03,0.05,0.1, 0.2$ and $0.3$ (from the bottom to the top). We have also depicted the recent BICEP2 data~\cite{bicep2} as well as the $\Lambda$CDM prediction (illustrated by the dashed curve). Note that there is a suppression of the $B$-mode power spectrum with respect to the predictions for a $\Lambda$CDM scenario. The suppression grows as the scale factor $a_{\textrm{f(R)}}$ decreases, that is, the earlier modified gravity starts operating, the larger the suppression in the tensor spectrum. This reduction in the $B$-mode amplitude is independent of the Hu $\&$ Sawicki model parameters $f_{R0}$ and $n$. This behaviour is due to the fact that $f_{R} \gg 1$ and both $f_{R}$, $f_{RR}$ are proportional to $|f_{R0}|$, making \eq{eq:tensor} behave as an over-damped harmonic oscillator. The depletion of the gravitational wave spectrum appears at small multipoles $\ell <10$ (see Fig.~\ref{fig:tensors2}) \ie at late times, where modifications of gravity start to be relevant. However, this multipole interval is clearly dominated by cosmic variance,  making the task of 
distinguishing $f(R)$ scenarios from $\Lambda$CDM through their $B$-mode signature extremely difficult and challenging. This conclusion might be avoided if the modifications of gravity start to be dominant at earlier times $z \gtrsim 20$, in which case structure formation measurements would put severe constraints on their viability.

\begin{figure*}
\begin{tabular}{c c}
\hspace{-1.cm}  \includegraphics[width=0.52\textwidth, height=7.1cm]{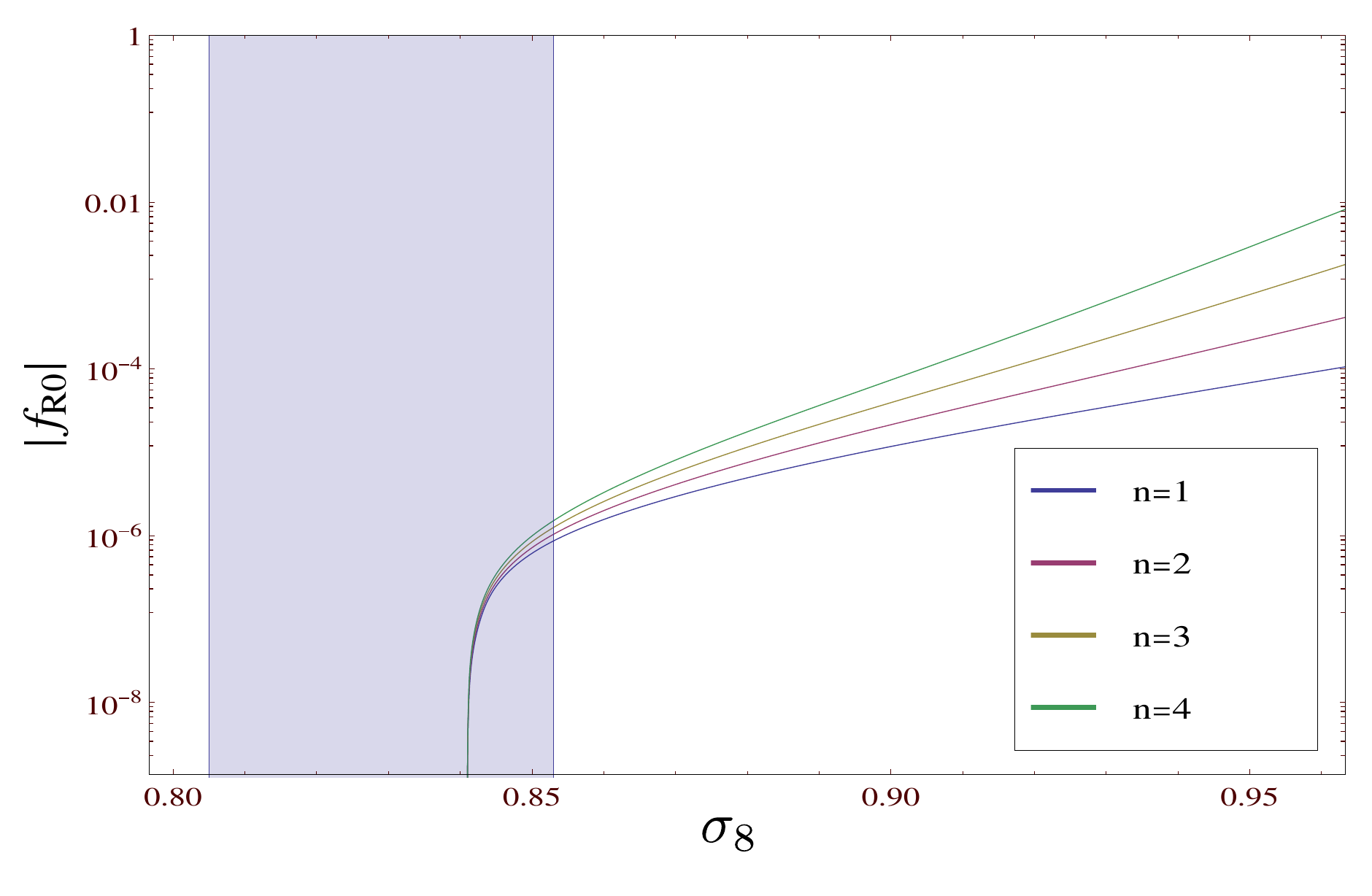}&
\includegraphics[width=0.5\textwidth, height=7cm ]
{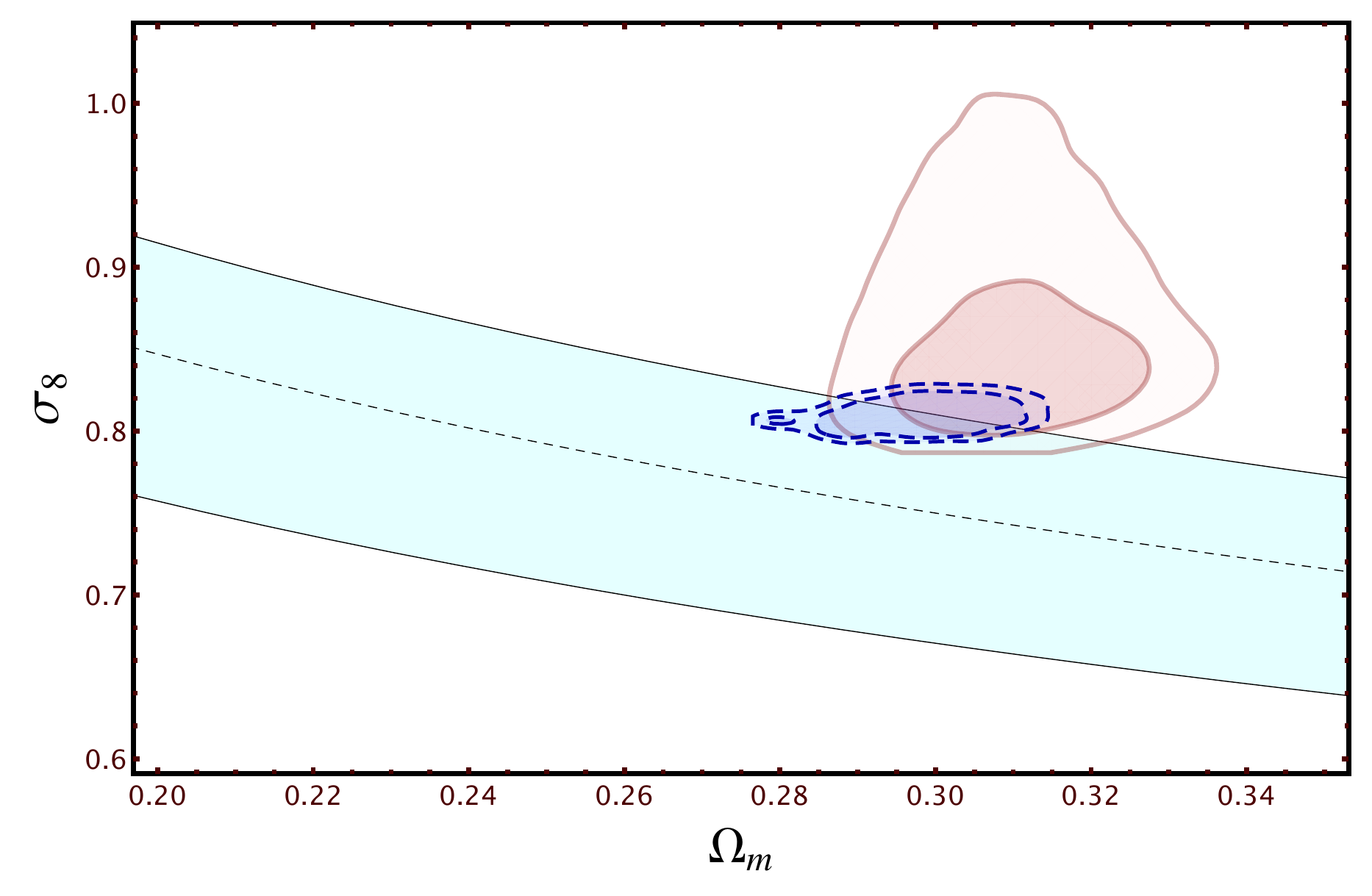}\\
\end{tabular}
 \caption{Left panel: values of the $\sigma_8$ clustering parameter as a function of the Hu $\&$ Sawicki parameters $|f_{R0}|$ and $n$. The  light blue (shaded) region denotes  the $2\sigma$ region from Planck measurements. Right panel: The shaded cyan region represents the $3\sigma$ constraints from the PSZ catalogue measurement of the $\sigma_8$-$\Omega_m$ relationship. The largest solid red contours depict the $68\%$ and $95\%$~CL constraints  arising from Planck temperature data in the context of the Hu $\&$ Sawicki $f(R)$ scenario, while the smallest dashed blue regions show the equivalent after combining Planck temperature with PSZ cluster measurements.}
\label{fig:sigma8}
\end{figure*}

\subsection{The $\sigma_8$-$\Omega_m$ degeneracy} 

\subsubsection{Galaxy Clusters}

Galaxy clusters are by far the largest virialised objects in the universe, and therefore they provide a unique way to probe the cosmological parameters. Cluster surveys usually measure the cluster number count  function $d N/d z$;  the number of clusters of a certain mass $M$ within a redshift interval (bin) $z+\delta z$, which, for a given survey, reads
\be
{d N\over dz}\Big|_{M>M_{\rm min}}=f_{\rm sky} {dV(z)\over dz}\int_{M_{\rm min}}^\infty dM \,{dn\over dM}(M, z)~,
\ee
where 
$f_{\rm sky}=\Delta\Omega/4\pi$ is the fraction of sky covered by the survey and 
\be
{dV(z)\over dz}={4\pi\over H(z)} \int_0^z dz' \({1\over H(z')}\)^2~.
\ee
 The cluster number count function is related to the underlying cosmological parameters, resulting in useful constraints. While the redshift is relatively easy to measure, the main uncertainty in this procedure comes from the cluster mass,  determined  through four main available methods: X-rays, velocity dispersion, SZ effect, and weak lensing.  The overall error in the cluster mass determination is usually around  $\Delta M/M\sim 10\%$.  Moreover, in order to relate the cluster number count function to the cosmological parameters, one has to input a  mass function $d n(z, M)/d M$ describing the abundance of virialised objects at a given redshift. This mass function is obtained through $N$-body simulations (see e.g. Ref.~\cite{Tinker:2008ff}), and depends on both the matter mass-energy density and on the standard deviation (computed in linear perturbation theory) of the density perturbations 
\be\label{eq:sigma8}
\sigma^2=\frac{1}{2\pi^2}\int^\infty_{0} dk k^2 P(k) W^{2}(kR)~,
\ee
\noindent where $P(k)$ is the matter power spectrum, $W(kR)$ the top-hat window function, 
\be\label{eq:wf}
W(kR)=3 \left(\sin(kR) - (kR) \cos(kR) \right)~, 
\ee
\noindent and $R$ is the comoving fluctuation size connected to the mass scale $M$ as $R = (3M/4\pi \rho_m)^{1/3}$. Taking into account these inherent uncertainties, there are still some degeneracies in the cosmological parameters probed by cluster surveys. The most-known one is the $\sigma_8$-$\Omega_m$ degeneracy, where $\Omega_m\equiv \rho_m/\rho_c$ is the current dark matter mass-energy density (normalised to the critical density) and $\sigma_8$ is the the root-mean-squared of density fluctuation in spheres of $8\,h^{-1}\mpc$ radius, see \eq{eq:sigma8}.

Therefore, the cosmological parameter constraints extracted from cluster number counts are usually reported by means of a relationship between the matter clustering amplitude $\sigma_8$ and the matter mass-energy density $\Omega_m$ parameters. More concretely, cluster catalogues provide the measurement of the so-called cluster normalisation condition, $\sigma_8 \Omega^\gamma_m$, where $\gamma \sim 0.4$~\cite{Allen:2011zs,Weinberg:2012es,Rozo:2013hha}. Here we will consider the cluster number counts as a function of the redshift from the Planck Sunyaev-Zeldovich (PSZ) catalogue~\cite{Ade:2013lmv}, as we will carefully describe in Sec.~\ref{sec:data}. In modified gravity theories, the matter power spectrum $P(k)$, and consequently the $\sigma_8$ parameter, are both modified in a non-trivial way, due to the presence of additional contributions in the perturbed Einstein equations. The linear matter power spectrum for a generic $f(R)$ theory can be found in Ref.~\cite{Bean:2006up}, while the non-linear clustering in the case of the Hu $\&$ Sawicki scenario has been explored numerically, using $N$-body simulations~\cite{Schmidt:2008tn,Li:2012by,Cai:2013toa}, and analytically, via modifications of the HALOFIT~\footnote{HALOFIT~\cite{Smith:2002dz} provides a modelling on the non-linear regime of the matter power spectrum, and it has been revisited by the authors of Ref.~\cite{Bird:2011rb} to account for massive neutrinos.} function~\cite{Zhao:2013dza}. Figure~\ref{fig:sigma8}, left panel, shows the values of $\sigma_8$ for different values of the Hu $\&$ Sawicki  parameters $|f_{R0}|$ and $n$. We also show the $2\sigma$ allowed region from Planck temperature data for this parameter, $\sigma_8=0.829\pm 0.012$~\cite{planck}. The value of the $\sigma_8$ parameter shows a clear departure from the $1\sigma$ allowed region by Planck data. For a fixed value of the $n$ parameter, this departure increases with the value of $|f_{R0}|$. Furthermore, this departure is more significant for smaller values of $n\simeq 1$, as expected from the naive analytical analysis performed in Sec.~\ref{sec:HS}. The right panel of Fig.~\ref{fig:sigma8} shows, in the ($\Omega_m$, $\sigma_8$) plane, as a shaded region, the $3\sigma$ constraints from the PSZ catalogue measurement of the $\sigma_8$-$\Omega_m$ relationship, $\sigma_8 (\Omega_m/0.27)^{0.3}=0.764\pm 0.025$ (at $68\%$~CL)~\cite{Ade:2013lmv}. The largest solid red contours depict the $68\%$ and $95\%$~CL constraints in the ($\Omega_m$, $\sigma_8$) plane arising from Planck temperature data in the context of the Hu $\&$ Sawicki $f(R)$ scenario, while the smallest dashed blue regions show the $68\%$ and $95\%$~CL constraints in the same plane after combining Planck temperature and SZ cluster measurements, as we shall describe in the following section. Notice that the allowed parameter region by Planck temperature data is noticeably reduced when adding the PSZ cluster catalogue data, allowing only small values for the $\sigma_8$ parameter and therefore constraining $|f_{R0}|$ to be smaller than $\sim 10^{-6}$. These findings will be fully explained and described in Sec.~\ref{sec:results}. 

\begin{table}[t!]
\begin{center}
\begin{tabular}{c|c}
\hline\hline
 Parameter & Prior\\
\hline
$\Omega_{b}h^2$ & $0.005 \to 0.1$\\
$\Omega_{c}h^2$ & $0.001 \to 0.99$\\
$\Theta_s$ & $0.5 \to 10$\\
$\tau$ & $0.01 \to 0.8$\\
$n_{s}$ & $0.9 \to 1.1$\\
$\ln{(10^{10} A_{s})}$ & $2.7 \to 4$\\
$|f_{\textrm{R0}}|$  &  $10^{-6} \to 0.1$\\
$n$ &  $1 \to 10$\\
\hline\hline
\end{tabular}
\caption{Uniform priors for the cosmological parameters used in this work.}
\label{tab:fRpriors}
\end{center}
\end{table}

\subsubsection{Weak lensing tomography}
Weak lensing tomography probes the matter power spectrum $P(k)$ via the correlations induced in the observed shape of distant galaxies by large scale structure. The observed ellipticity of a distant galaxy $\epsilon_{obs}$ is related to its intrinsic ellipticity $\epsilon_s$ by means of the cosmic shear $\gamma$: $\epsilon_{obs}= \epsilon_{s} + \gamma$.  The 2-point shear correlation function is then obtained after averaging over all galaxy pairs distant by an angle $\theta$. Therefore, the galaxy sample is divided into redshift bins, cross-correlating the extracted shear after summing after over all possible galaxy pairs in the two redshift bins $i$ and $j$. This estimate needs to be compared with the theoretical prediction for the 2-point shear correlation function, which depends on the underlying cosmology  via the convergence power spectrum in the $i$, $j$ redshift bins
\begin{equation}
P^{ij}_\kappa (\ell)= \frac{9 H^4_0 \Omega^2_m}{4}\int^{\chi_h}_{0} d\chi^\prime \frac{g_i(\chi^\prime)g_j(\chi^\prime)}{\chi^2} P\left(\frac{\ell}{\chi},\chi^\prime\right)~,
\end{equation}
where $H_0$ is the current value of the Hubble parameter, $\chi^\prime$ is the comoving radial distance, $\chi$ is the comoving angular diameter distance and $\chi_h$ is the horizon distance. 
The factors $g_i(\chi^\prime)$ and $g_j(\chi^\prime)$ are the lensing weights, which depend on the normalised galaxy distribution in each of the redshift bins, see e.g. Refs.~\cite{Schrabback:2009ba,Heymans:2013fya}. Weak lensing measurements alone are only sensitive to the overall amplitude of the matter power spectrum $P(k)$, which in turn depends on a combination of the $\sigma_8$ parameter and the current matter mass-energy density $\Omega_m$. Tomographic lensing surveys provide therefore additional and independent constraints on the relationship  between $\sigma_8$ and  $\Omega_m$.

\section{Method and Cosmological datasets}
\label{sec:data}

The Hu $\&$ Sawicki $f(R)$ gravity model explored here is described by the following parameters:
\begin{equation}
\label{parameter}
  \{\omega_b,\omega_c, \Theta_s, \tau, n_s, \log[10^{10}A_{s}], |f_{R0}|, n\}~,
\end{equation}

\noindent where $\omega_b\equiv\Omega_bh^{2}$ and $\omega_c\equiv\Omega_ch^{2}$  
are the physical baryon and cold dark matter energy densities,
$\Theta_{s}$ is the ratio between the sound horizon at decoupling and the angular
diameter distance to the last scattering surface, $\tau$ the optical depth to reionisation,
$n_s$ is the scalar spectral index, $A_{s}$ the amplitude of the
primordial spectrum and the parameters $|f_{R0}|$ and $n$ are the Hu $\&$ Sawicki parameters, see Eqs.~(\ref{eq:fr})  and (\ref{eq:limit}).
The priors used for these parameters are specified in Tab.~\ref{tab:fRpriors}. 

For numerical purposes, we have used the modified version of the publicly available Boltzmann CAMB
code~\cite{camb} for modified gravity models, MGCAMB~\cite{Hojjati:2011ix,Zhao:2008bn}, extracting the cosmological parameters with the  Monte Carlo Markov Chain (MCMC) package \texttt{cosmomc}~\cite{Lewis:2002ah}. There have been studies in the literature showing that the so-called chameleon mechanism may affect the halo mass function~\cite{Schmidt:2008tn,Schmidt:2009am,Ferraro:2010gh,Li:2011uw,Kopp:2013lea,Lombriser:2013wta},  showing that for small values of $|f_{R0}|$ the value of the $\sigma_8$ parameter computed from CAMB using pure linear theory may be also inappropriate. The change in the $\sigma_8$ parameter introduced by the chameleon mechanism could be addressed via simulations or additional fitting functions to the linear matter power spectrum.  Indeed, the MGCAMB version that we exploit includes a modification of the HALOFIT function for the Hu $\&$ Sawicki model from Ref.~\cite{Zhao:2013dza}. Such a non-linear description provides an accuracy below $10\%$ provided that $|f_{R0}|\gtrsim 10^{-6}$. 
This justifies our choice of the lower prior in $|f_{R0}|$  of $10^{-6}$, as this is the lowest value for which MGHALOFIT has been calibrated. 

Our basic dataset is the  Planck CMB temperature anisotropies data, including also the lensing likelihood, see Refs.~\cite{Ade:2013ktc,Planck:2013kta}, together with the 9-year polarization data from the WMAP satellite~\cite{Bennett:2012fp}. The likelihood for the former datasets is computed by means of  the Planck collaboration publicly available likelihood tools, see Ref.~\cite{Planck:2013kta} for more details.

\begin{table*}[!t]
\begin{center}
\begin{tabular}{p{1cm} lccccc}
\hline \hline
\hspace{1mm}\\
                        & CMB &     CMB+BAO & CMB+BAO& CMB+BAO & CMB+BAO\\
                        &          &              &          +SZ Clusters$^1$   &   +SZ Clusters$^2$ &+CFHTLens \\                
\hline
\hspace{1mm}\\
               
$|f_{\textrm{R0}}|$ & $<4.8 \times 10^{-4}$ & $<8.3\times 10^{-4}$& $<3.7\times 10^{-6}$ & $<1.6\times 10^{-5}$ & $<6.5\times10^{-5}$\\
\hspace{0.5mm}\\
$\Omega_{\textrm{m}}$ &$0.37_{-0.04}^{+0.04}$ &$0.32_{-0.02}^{+0.02}$  &$0.291_{-0.002}^{+0.006}$ &$0.31_{-0.02}^{+0.02}$&$0.30_{-0.02}^{+0.02}$\\
\hspace{0.5mm}\\
$\sigma_{8}$ &$0.87_{-0.04}^{+0.06}$ &$0.86_{-0.05}^{+0.06}$&$0.798_{-0.002}^{+0.003}$ &$0.81_{-0.01}^{+0.02}$&$0.82_{-0.02}^{+0.02}$\\[1mm]
\hline
\hline
\end{tabular}
\caption{$68\%$~CL errors on the $\sigma_8$ and the $\Omega_m$ parameters, as well as the upper $95\%$~CL limits on the $|f_{R0}|$ parameter of the Hu $\&$ Sawicki model for the different possible data combinations considered in the analysis.}
\label{tab:limits}
\end{center}
\end{table*}

The BICEP2 collaboration, after three years of collecting data, has recently reported  evidence for the detection of $B$-modes in the multipole range $30 <\ell< 150$~\cite{bicep2,bicep22}, with $6\sigma$ significance. However, as we have seen before, the effects of the modified gravity model studied here on the tensor spectrum appear in the $\ell<10$ multipole range, region not covered by the BICEP2 data, where cosmic variance dominates. Therefore, we shall neglect the analysis of tensor modes in the following. 

We have also considered Baryon Acoustic Oscillations (BAO) data, which are the imprint of the competition between gravity and radiation pressure in the coupled baryon-photon fluid before the recombination era. The BAO data considered here include the 6dF galaxy survey measurements at a redshift $z=0.106$~\cite{Beutler:2011hx} and the WiggleZ Survey BAO measurements at $z=0.44, 0.6$ and $0.73$~\cite{Blake:2011en}. These former measurements refer to the spherically averaged clustering statistics, which is a  function of the angular diameter distance and the Hubble expansion rate at a given redshift. However, if measurements of the BAO signature along and across the line of sight are feasible, a separate measurement of the Hubble parameter and the angular diameter distance are possible. The Data Release 11 (DR11) from the Baryon Oscillation Spectroscopic Survey (BOSS)~\cite{Dawson:2012va} experiment, part of the Sloan Digital Sky Survey III (SDSSIII) program~\cite{Eisenstein:2011sa}, provides, at an effective redshift of $0.57$, $D_A(z=0.57)=1421\pm 20$ Mpc\,$\times \(r_s(z_{drag})/r_{s,fid}\) $ and $H(z=0.57)= 96.8\pm 3.4$ km/s/Mpc $\times (r_s(z_{drag})/r_{s,fid})$~\cite{Anderson:2013zyy}, where $r_{s,fid}=149.28$~Mpc is the sound horizon at the recombination period in the fiducial model and $z_{drag}$ is usually defined as the epoch in which the drag optical depth equals $1$. We have also exploited the BOSS Lyman alpha forest BAO signature~\cite{Font-Ribera:2014wya}. The corresponding effective redshift is $z=2.36$, and the constraints are $c/(H(z=2.36) r_s( z_{drag})=9.0\pm 0.3$ and $D_A(z=2.36)/ r_s( z_{drag})=10.8 \pm 0.4$~Mpc. 

As previously discussed, independent constraints on the  power spectrum amplitude $\sigma_8$ and on the dark matter mass-energy density $\Omega_m$ are obtained from the abundance of clusters as a function of the redshift. We shall consider here the cluster number counts as a function of the redshift from the Planck Sunyaev-Zeldovich (PSZ) catalogue, the largest SZ cluster sample with 189 galaxy clusters~\cite{Ade:2013lmv}. The relationship between the matter clustering amplitude and the matter mass-energy density is $\sigma_8 (\Omega_m/0.27)^{0.3}=0.764\pm 0.025$. Since the value of the $\sigma_8$ parameter is degenerate with the cluster mass bias, fixing the bias parameter to the value obtained from numerical simulations improves considerably the error on the $\sigma_8$-$\Omega_m$ relationship: $\sigma_8 (\Omega_m/0.27)^{0.3}=0.78\pm 0.01$. We shall refer to the optimistic and pessimistic case as $(1)$ and $(2)$, respectively. 

\begin{figure*}[!t]
\begin{tabular}{c c}
\includegraphics[width=8.2cm]{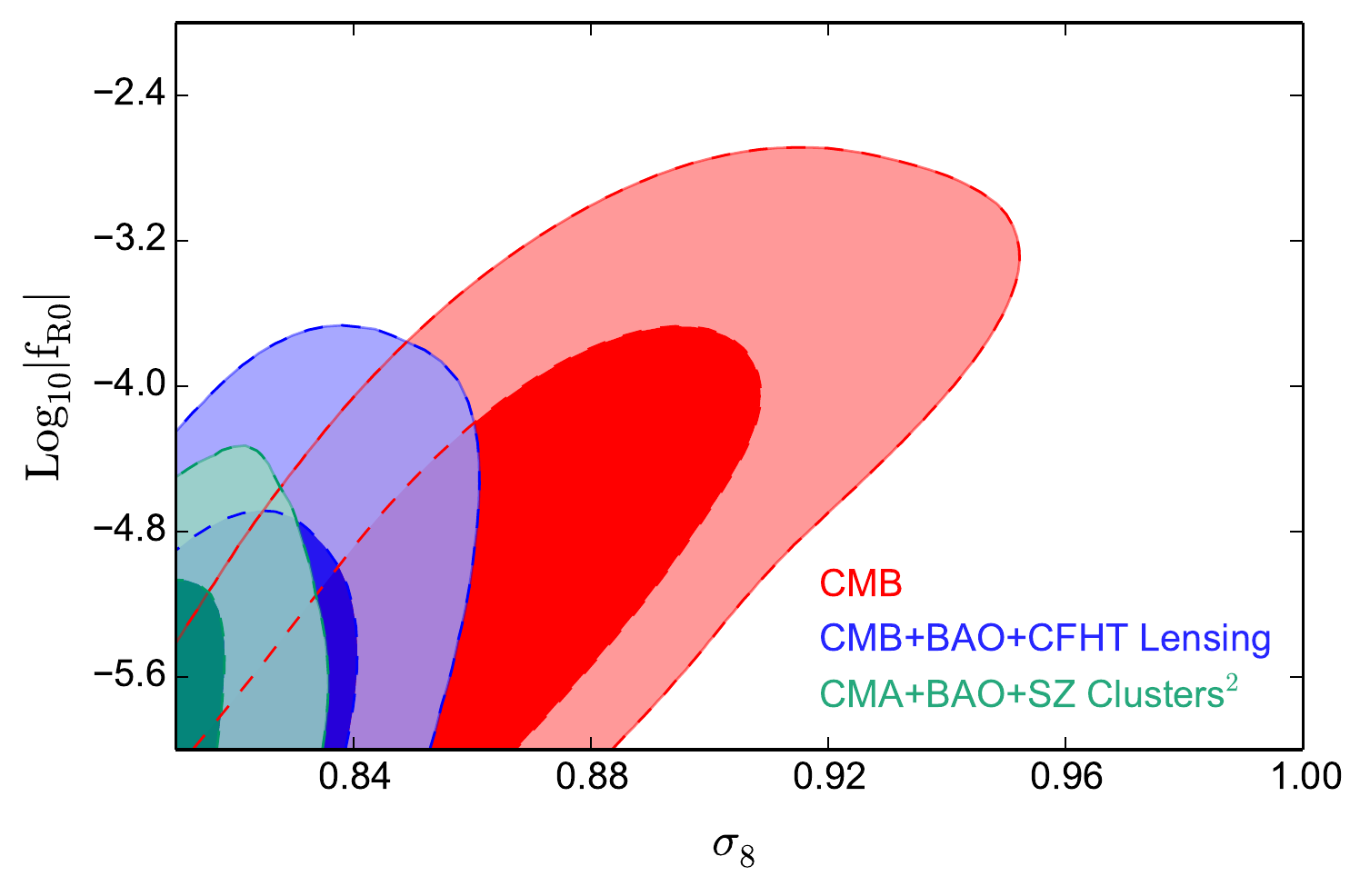}&\includegraphics[width=8.4cm]{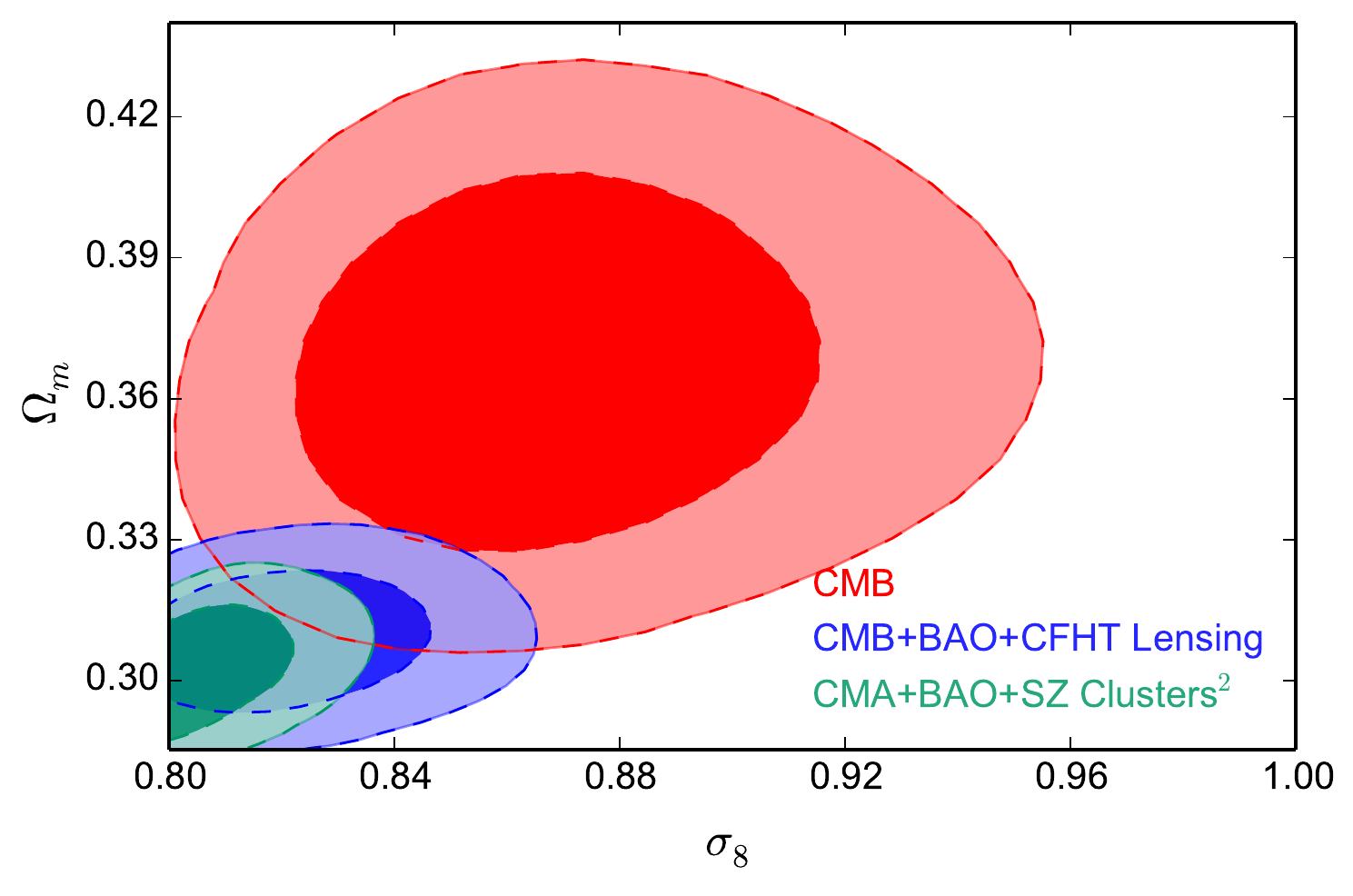}\\
\end{tabular}
 \caption{The left panel depicts the $68\%$ and $95\%$~CL allowed regions in the ($\sigma_8$, $|f_{R0}|$) plane for different possible data combinations. In the case of the PSZ catalogue, we have only illustrated the most conservative case. The right panel shows the equivalent but in the ($\sigma_8$, $\Omega_m$) plane. }
\label{fig:mcmc2}
\end{figure*}

On the other hand, other independent constraints on the relationship  between $\sigma_8$ and  $\Omega_m$ arise from tomographic weak lensing surveys via measurements of the galaxy power shear spectra. The CFHTLens survey finds $\sigma_8 (\Omega_m/0.27)^{0.46}=0.774^{+0.032}_{-0.041}$~\cite{Heymans:2013fya}, after analysing six tomographic redshift bins ranging from $z=0.28$ to $z=1.12$. 

The relationship between  $\sigma_8$ and  $\Omega_m$ from both the PSZ clusters catalogue and the CFHTLens experiment are added in our analyses by applying these constraints to our Monte Carlo Markov chains. In order to do so, we perform a post-processing of the MCMC chains obtained from the analysis of the CMB and BAO data described above. 

We conclude this section by commenting on the compatibility of the datasets used in the present analysis. The $\sigma_8$ measurements from cosmic shear and cluster counts show both a $2\sigma$ discrepancy with Planck CMB temperature power spectrum estimates (see Sec. 5.5 of Ref.~\cite{planck}). This tension could be due to biases in the calibration of cluster masses, and including extra parameters in the minimal $\Lambda$CDM model may help in reconciling  these two datasets~\cite{Hamann:2013iba,Giusarma:2014zza,Dvorkin:2014lea}.  In the present analysis, we will follow as well this quite non-conservative approach, using this additional independent $\sigma_8$-measurements to further constrain the Hu $\&$ Sawicki model parameters. 

\section{Results}
\label{sec:results}

We show in Table~\ref{tab:limits} the $68\%$~CL errors on the $\sigma_8$ and the $\Omega_m$ parameters, as well as the upper $95\%$~CL limits on the $|f_{R0}|$ Hu $\&$ Sawicki parameter, for the different possible data combinations explored in this study. Notice that we do not show the limits on the $n$ parameter, as both the $95\%$~CL upper and lower limits coincide  with the upper and lower priors respectively. This is related to the fact that the effect of $n$ on the cosmological observables is much milder than the effect induced by $f_{R0}$. Indeed, while smaller values of $|f_{R0}|$ will produce cosmological histories closer to a $\Lambda$CDM expansion model, a larger value of $n$ will mimic the $\Lambda$CDM scenario until later in the expansion history~\cite{Hu:2007nk}. However, modifications of gravity start to be important at relatively low redshift, and therefore the impact of $n$ is not as significant as the one of $f_{R0}$.

The $95\%$~CL limit we get on $|f_{R0}|$ considering CMB data (with lensing included) is $|f_{R0}|< 4.8 \times 10^{-4}$. If on the other hand the CMB lensing information is neglected, the former bound weakens by one order  of magnitude. Similar conclusions concerning the impact of CMB lensing on cosmological analyses of $f(R)$ theories have been found in Ref.~\cite{Hu:2013aqa}. The above $95\%$~CL constraint of $|f_{R0}|< 4.8 \times 10^{-4}$ quoted above is similar to the recent limit $|f_{R0}|< 3\times 10^{-6}$ (at $68\%$~CL)~\cite{Bel:2014awa} and it is milder than the one found using the WiggleZ survey~\cite{Dossett:2014oia}, $|f_{R0}|< 1.4\times 10^{-5}$ at  $95\%$~CL. 

However, once we add weak lensing tomography information on the $\sigma_8$-$\Omega_m$ relationship  from CFHTLens, we get $|f_{R0}|< 6.5\times 10^{-5}$ at $95\%$~CL. Adopting the pessimistic case of the PSZ catalogue, the $95\%$~CL upper limit on $|f_{R0}|$ is $1.6\times 10^{-5}$, which is comparable to the limit from the WiggleZ survey quoted above. On the other hand, the PSZ optimistic scenario provides a bound $|f_{R0}|< 3.7\times 10^{-6}$ at  $95\%$~CL. This bound is similar to the limit  $|f_{R0}|< 3\times 10^{-6}$ at $95\%$~CL found  in Ref.~\cite{Smith:2009fn} studying strong lensing galaxy signals.  However, our more optimistic case ($|f_{R0}|<3.7\times 10^{-6}$ at $95\%$~CL) is milder than the limits arising from galactic physics, either from distance indicators ($|f_{R0}|<5\times 10^{-7}$ at $95\%$~CL~\cite{Jain:2012tn}) or from the fifth-force effects on diffuse dwarf galaxy components~\cite{Vikram:2013uba,Vikram:2014uza}.

Table~\ref{tab:limits} also shows that, when considering additional cluster SZ data or weak lensing tomography measurements, the mean values of $\sigma_8$ are shifted towards smaller values, and the errors on this parameter are reduced roughly by a half. 

Figure~\ref{fig:mcmc2}, left panel, depicts the $68\%$ and $95\%$~CL allowed regions in the ($\sigma_8$, $|f_{R0}|$) plane for different possible data combinations. Notice that the regions arising from the analysis to Planck and BAO data show a degeneracy which is identical to that shown in the left panel of Fig.~\ref{fig:sigma8}, indicating that $|f_{R0}|$ is positively correlated with $\sigma_8$. However, once that additional constraints on the $\sigma_8$-$\Omega_m$ relationship are included in the analysis, the degeneracy is broken and larger values of $|f_{R0}|> 10^{-6}$ are highly disfavoured. The right panel of Fig.~\ref{fig:mcmc2} shows the $68\%$ and $95\%$~CL allowed regions in the ($\sigma_8$, $\Omega_m$) plane. Considering CMB and BAO data only, these two parameters show a quite large correlation, because of their very similar effects on the power spectrum normalisation, as previously observed from the mean values and errors quoted in Tab.~\ref{tab:limits}. When independent constraints on the relationship between these two parameters are also considered, as those from cosmic shear and/or counts of rich clusters, the correlation among them is highly reduced, as well as the errors on the clustering parameter $\sigma_8$.

\section{Conclusions}
\label{sec:conclusions}
Cosmological measurements of the Cosmic Microwave Background (CMB), of type Ia Supernovae and of the large scale structure in the universe have robustly established that the universe is currently expanding at an accelerating rate. The nature of the physics responsible for such a phenomenon remains still obscure. A possible explanation is that cosmic expansion is due to modifications in the gravitational sector  at very large scales. Among the plethora of possible models, in this work, we explore the observational signatures of the Hu $\&$ Sawicki model, which satisfies solar system constraints, and depends on two parameters, $f_{R0}$ and $n$. The effects of such a model in the Integrated Sachs Wolfe effect and in the CMB lensing have been carefully explored  by some authors. However, a complete study including both tensor modes and additional constraints on the power spectrum amplitude was lacking in the literature. Here we address these issues, computing from first principles the modified gravitational wave spectrum in this model. Even if there is a modification in the damping and source terms in the tensors equations (partially induced by the non-zero anisotropic stress inherent to these family of models), the effects of the modified gravity model studied here on the tensor spectrum appear only in the $\ell<10$ multipole range. Given that this region is not covered by the BICEP2 experiment and furthermore it is cosmic variance dominated, we conclude that tensor modes can not help in distinguishing modified gravity from dark energy scenarios.  Hence, we have neglected tensor modes in our numerical simulations. 
We have carried out a Markov Chain Monte Carlo (MCMC) analysis, using the most recent CMB and Baryon Acoustic Oscillation (BAO) datasets as well as independent constraints on the relationship between the matter clustering amplitude $\sigma_8$ and the matter mass-energy density $\Omega_m$ from Planck Sunyaev Zeldovich (PSZ) cluster number counts,  as well as from the CFHTLens weak lensing tomography measurements.  Combining CMB, BAO and the most optimistic  $\sigma_8$-$\Omega_m$ relationship from the PSZ catalogue, we obtain a bound on the parameter  $|f_{R0}|<3.7\times 10^{-6}$ at $95\%$~CL. This constraint is competitive with strong galaxy lensing limits and stronger than previous estimates considering cosmological data only~\cite{Dossett:2014oia,Bel:2014awa}. If we instead consider the more pessimistic approach for the PSZ catalogue, in which the cluster mass bias is taken as a free parameter, we obtain $|f_{R0}|<1.6\times 10^{-5}$ at $95\%$~CL. Concerning the other parameter describing the model, $n$, it turns out to be unconstrained, as its impact on the different cosmological observables is much milder than the impact of the $f_{R0}$ parameter. Future cluster surveys covering 10,000 deg$^2$ in the sky, with galaxy surface densities of  $10$~arcmin$^{-2}$ ($30$~arcmin$^{-2}$), combined with forthcoming weak lensing data, could reach a precision in the $\sigma_8$-$\Omega_m$ relationship of $\lesssim 1\%$ ($\lesssim 0.06\%$)~\cite{Weinberg:2012es}, whereas $1.3\%$ is the more optimistic error that we have considered here with current data. This expected reduced uncertainty in the $\sigma_8$-$\Omega_m$ relationship could improve further the bounds obtained in the present analysis, provided that N-body simulations for small values of $|f_{R0}|$ (i.e. $|f_{R0}|<\mathcal{O}(10^{-6})$) become available in the literature.

\acknowledgments
O.M. is supported by the Consolider Ingenio project CSD2007-00060, by
PROMETEO/2009/116, by the Spanish Ministry Science project FPA2011-29678 and by the ITN Invisibles PITN-GA-2011-289442. We also thank the Spanish MINECO (Centro de excelencia Severo Ochoa Program) under grant SEV-2012-0249.

\end{document}